\documentclass[useAMS,usenatbib]{mn2e}

\usepackage{epsfig, natbib, graphicx, color}

\input epsf 

\usepackage{color}

\def\lesssim{\mathrel{\hbox{\rlap{\hbox{\lower3pt\hbox{$\sim$}}}\hbox{\raise2pt\hbox{$<$}}}}}
\def\gtrsim{\mathrel{\hbox{\rlap{\hbox{\lower3pt\hbox{$\sim$}}}\hbox{\raise2pt\hbox{$>$}}}}}

\newcommand{\Mpc}{\mbox{Mpc}}

\newcommand{\msun}{M_\odot}

\newcommand{\be}{\begin{equation}}
\newcommand{\ee}{\end{equation}}
\newcommand{\bea}{\begin{eqnarray}}
\newcommand{\eea}{\end{eqnarray}}

\newcommand{\Ysz}{Y_{\rm SZ}}

\newcommand{\eV}{\mbox{eV}}

\title{Cluster Cosmology at a Crossroads: Neutrino Masses}

\author[E. Rozo, E. S. Rykoff, J. G. Bartlett, A. Evrard]{E. Rozo$^{1}$, E. S. Rykoff$^{1}$,
J.G. Bartlett$^{2,3}$, A. Evrard$^{4}$.\\
$^{1}${SLAC National Accelerator Laboratory, Menlo Park, CA 94025.}\\
$^{2}${APC, AstroParticule et Cosmologie, Universit\'e Paris Diderot, CNRS/IN2P3, CEA/lrfu, Observatoire de Paris,}\\
{\hspace{5mm} Sorbonne Paris Cit\'e, 10, rue Alice Domon et L\'eonie Duquet, Paris Cedex 13, France.}\\
$^{3}${Jet Propulsion Laboratory, California Institute of Technology, 4800 Oak Grove Drive, Pasadena, CA, U.S.A.}\\
$^{4}${Departments of Physics and Astronomy and Michigan Center for Theoretical Physics, University of Michigan, Ann Arbor, MI 48109.}
}

\begin{document}

\maketitle

\label{firstpage}

\begin{abstract}
Galaxy clusters --- in combination with CMB and BAO data --- can provide precise constraints on the sum of neutrino
masses.  However, these constraints depend on the calibration of 
the mass--observable relation.  For instance, 
the mass calibration employed in \citet{planck11_xray,planck11_local} rules out the minimal 6-parameter
$\Lambda$CDM model at $3.7\sigma$, and implies a sum of neutrino masses $\sum m_\nu = 0.39 \pm 0.10$.
By contrast, the mass calibration favored by \citet{rozoetal12d} from a self-consistent analysis of X-ray, SZ, and optical
scaling relations is consistent with a minimal flat $\Lambda$CDM model with no massive neutrinos ($1.7\sigma$),
and is a better fit to additional data (e.g. $H_0$).   We discuss these results in light of the most recent SPT and ACT
analyses, and the implications of our results on the current mild
``tension'' ($<2\sigma$) between CMB and BAO+$H_0$ data.
\end{abstract}


\section{Introduction}

Galaxy clusters provide an important complementary probe to the Cosmic Microwave Background (BAO) 
and geometric probes like Baryon Acoustic Oscillations (BAO).  
Specifically, clusters provide precise estimates of the so called cluster normalization
condition, $\sigma_8\Omega_m^\gamma$, where $\gamma\approx 0.5$ \citep[see][for a general review
of cluster cosmology]{allenetal11,weinbergetal12}.
While the CMB provides an accurate
measurement of the amplitude of the power spectrum at the epoch of last scattering, the corresponding
constraint on the cluster normalization condition can be highly uncertain due to the extrapolation from 
$z\approx 1200$ to $z\approx 0$.  This uncertainty is primarily dominated by the impact of $\Omega_m$
on the growth function,  but also depends on additional
cosmological parameters such as curvature, the dark energy equation-of-state, and, most relevant
for our purposes, neutrino masses.
By directly measuring the amplitude of matter fluctuations in the low redshift universe
and comparing to the range of theoretical predictions from CMB+BAO data, galaxy
clusters allow us to
improving cosmological constraints on these parameters over and above 
the CMB+BAO only results \citep[e.g.][]{bureninvikhlinin12,mantzetal10c,reidetal10}.

However, the cosmological constraints from galaxy clusters are critically dependent on our 
ability to estimate cluster masses.
We illustrate this basic argument using the results of \citet[][hereafter V09]{vikhlininetal09b}, 
highlighting how the constraints on neutrino mass depend on the adopted cluster mass calibration.
In particular, we consider two additional mass calibrations, that adopted in \citet{planck11_xray,planck11_local}
and that of \citet{rozoetal12d}.  We then connect these arguments to the recent results
from SPT \citep{houetal12} and ACT \citep{sieversetal13,hasselfieldetal13}, with some emphasis 
on the mild tension ($< 2\sigma$) between CMB and BAO data in the current analyses.

All of our results are computed using importance sampling of the WMAP9 chains \citep{wmap9}.
The likelihood distributions are computed using a Kernel Density Estimator (KDE), where each point
is assigned the weight reported in the WMAP9 chains.
When adding galaxy clusters, we rely on the fact that low-redshift galaxy clusters only constrain
a specific combination of cosmological parameters, namely $\sigma_8\Omega^\gamma$ where
$\gamma\approx 0.5$, with this constraint being essentially independent of the remaining
cosmological parameters.  Consequently, galaxy clusters modify the weight $w_i$ for 
each point in the WMAP9 chains via
\be
w_{i,cl} = w_i \exp \left[ - \frac{1}{2} \frac{(s_{8,i} - s_{8,prior})^2}{\sigma_s^2} \right]
\ee
where $w_i$ is the original weight, and $s_8=\sigma_8\Omega_m^\gamma$
is the relevant cluster normalization condition.  
All upper limits on neutrino masses are 95\% confidence.  Constraints of the form
$\sum m_\nu = X^{+a}_{-b}$ imply $X$ is the maximum likelihood point, and
$a$ and $b$ define the $68\%$ confidence contour.  All constraints are reported
after marginalizing over the remaining model parameters.
For a general review of cosmological bounds on neutrino masses, we refer the reader to \citet{lesgourguespastor12}.
Recent reviews on cluster cosmology can be found in \citet{weinbergetal12,allenetal11}.

Throughout, a minimal $\Lambda$CDM model references a flat $\Lambda$CDM models with 
only 6 free parameters: the amplitude of the primordial power spectrum fluctuations $A_s$, 
the tilt of the primordial power spectrum $n_s$, the matter density $\Omega_m$, the hubble
parameter $h$, the angular scale of the sound horizon at last scattering $\theta_s$,
and the optical depth to the surface of last scattering $\tau$.


\section{How Galaxy Clusters Interact with CMB+BAO Constraints}

V09 provides a precise constraint on 
the quantity $s_8 \equiv \sigma_8(\Omega_m/0.25)^{0.47}=0.813 \pm 0.013$.  The error bar
is statistical errors only: in the spirit of the V09 analysis, 
we consider systematic shifts in the mass scale independently.
By comparison,  the uncertainty from WMAP9 data only \citep{wmap9}
in a minimal $\Lambda$CDM model 
is significantly larger, $s_8 =  0.866 \pm 0.058$. Note that because $s_8$ was defined
using $\Omega_m=0.25$ as a reference value, the fact that $\Omega_m\approx 0.28$
from WMAP9 data implies that the quoted $s_8$ value is significantly higher
than the $\sigma_8$ value derived from WMAP9, $\sigma_8=0.82$.
The uncertainty in $s_8$ can be reduced with an independent
probe of $\Omega_m$, or, since WMAP9 constrains $\Omega_m h^2$, a measurement of $h$.
The most significant improvement occurs when one adds BAO 
information \citep[based on the analyses in][]{beutleretal11,padmanabhanetal12,andersonetal12,blakeetal12},
which results in $s_8=0.898 \pm 0.029$ after marginalizing over all other parameters.  
This value is borderline consistent with the V09 result ($2.7\sigma$).

One can decrease the modest tension between WMAP9+BAO and V09 by allowing for dynamical dark energy
or curvature.  However, the single extension that leads to the largest improvement is allowing for non-zero
neutrino masses.  Because neutrinos can escape their initial density peak, massive neutrinos effectively smear out a fraction
of the mass over the neutrino free streaming scale, leading to a reduced value of the predicted amplitude of matter
fluctuations at $z=0$.   In this extension of the minimal flat $\Lambda$CDM model, the WMAP9+BAO prediction for $s_8$
after marginalizing over all other parameters is $s_8= 0.828 \pm  0.053$, in excellent agreement with the V09 result ($0.2\sigma$ offset).

\begin{figure}
\hspace{-5mm} \includegraphics[width=90mm]{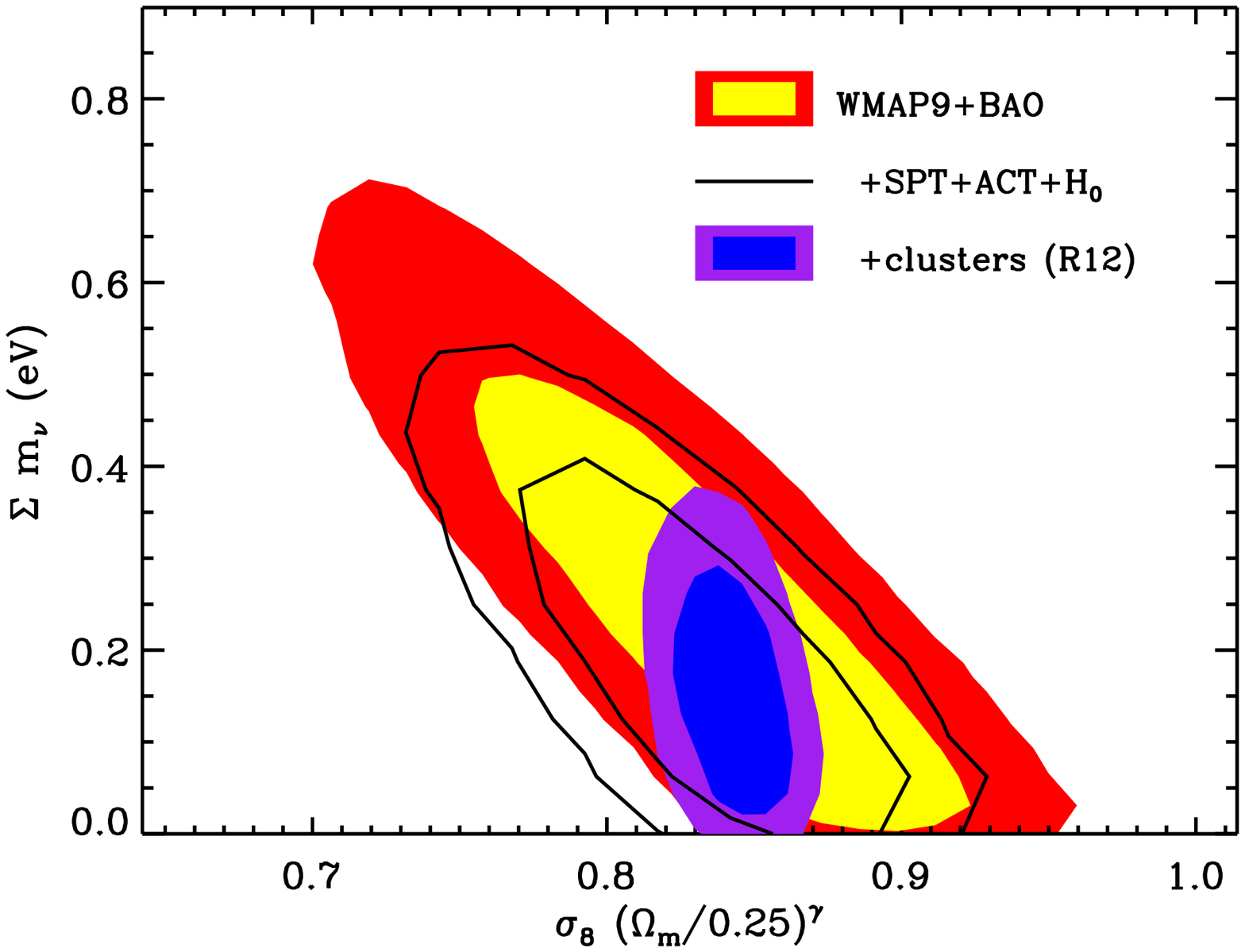}
\caption{68\% and 95\% confidence contours in the $s_8$--$\sum m_\nu$ plane,
where $s_8=\sigma_8(\Omega_m/0.25)^{0.47}$.
Red/yellow ellipses are from WMAP9+BAO data only, while the solid curves
add SPT, ACT, and $H_0$
data as in \citet{wmap9}.  
The blue/purple contours show
the impact of adding galaxy clusters with the \citet{rozoetal12d} mass calibration.
}
\vspace{-3mm}
\label{fig:lkhd2d}
\end{figure}

The WMAP9+BAO only constraint on the sum of neutrino masses is $\sum m_\nu \leq 0.58$.  As shown
by the red and yellow contours in Figure \ref{fig:lkhd2d}, this constraint
is strongly degenerate with the cluster 
normalization condition, a degeneracy that persists even if one adds SPT, ACT, and $H_0$ data (solid line ellipses).
The origin of this degeneracy is clear: the CMB constrains the amplitude of matter fluctuations at last scattering.
Massive neutrinos take part of the initial mass fluctuations and ``spread it out'' over the free-streaming scale, leading
to a smoother Universe (lower $\sigma_8$).  The more mass one spreads --- i.e. the more massive the neutrinos are ---
the smoother the late Universe is, leading to the observed anti-correlation.

Adding galaxy clusters breaks this degeneracy.  Using the V09 constraints we find
\be
\sum m_\nu = 0.31 ^{+0.10}_{-0.11}\ \eV\ \hspace{0.2in} \mbox{WMAP9+BAO+Cl (V09)}
\ee
Note that the purple/blue contours in Figure \ref{fig:lkhd2d} are not those obtained with the original V09
cluster normalization condition, reflecting instead the results obtained using the \citet{rozoetal12d} mass calibration.
We discuss this result in more detail below. 
For a more detailed discussion of the V09 results using WMAP7 data and BAO constraints pre-BOSS, 
see \citet{bureninvikhlinin12}.


\section{The Role of Mass Calibration}

The sensitivity of the cluster normalization condition to the mass calibration of galaxy clusters is
intuitively obvious: as one increases the mass assigned to galaxy clusters, the resulting cosmological constraints
result in a more inhomogeneous Universe (higher $\sigma_8$) with higher matter density.  Thus,
higher cluster masses result in higher cluster normalization conditions.
Because of the strong $\sum m_\nu$--$s_8$ degeneracy in the WMAP9+BAO data,
it follows that cluster mass calibration can have a dramatic impact on the recovered neutrino mass.

We illustrate the importance of cluster masses on neutrino mass constraints by considering how the
WMAP9+BAO+Cl constraint change as we shift the mass calibration away from that employed in V09.
To compute the cluster normalization condition for an arbitrary mass calibration, we rely on V09,
who shows that systematically shifting the mass of all galaxy clusters by $\pm 9\%$
shifts the corresponding $s_8$ value by $\pm 0.024$.
Consequently, for small mass shifts $\Delta \ln M$ is the mass calibration offset relative to V09, 
we can approximate the $s_8$ dependence on cluster mass calibration via
\be
s_8 = s_{8, V09} + 0.024 \frac{ \Delta \ln M}{0.09}.
\ee
\vspace{0.05in}


\begin{figure*}
\includegraphics[width=85mm]{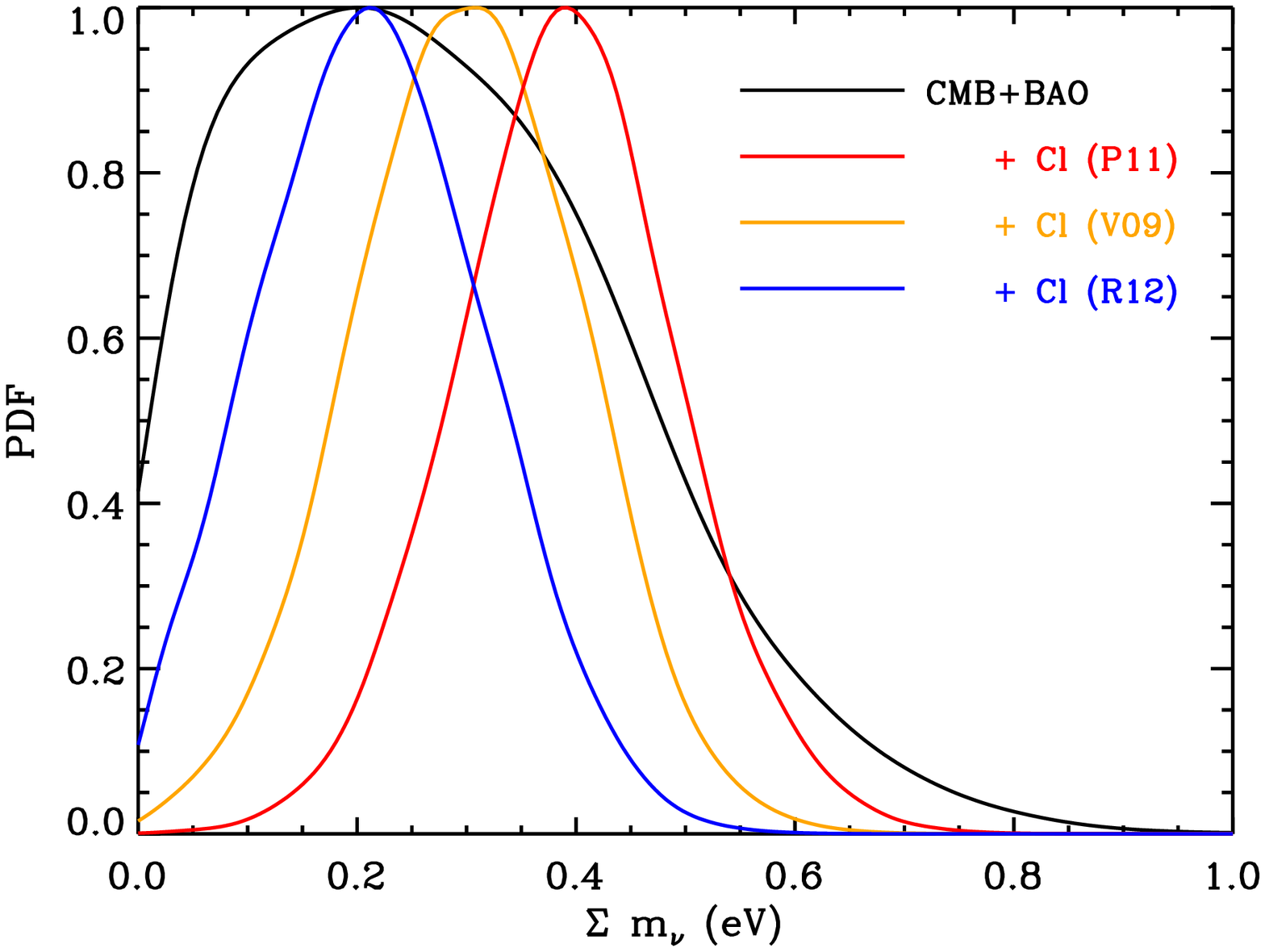}
\includegraphics[width=85mm]{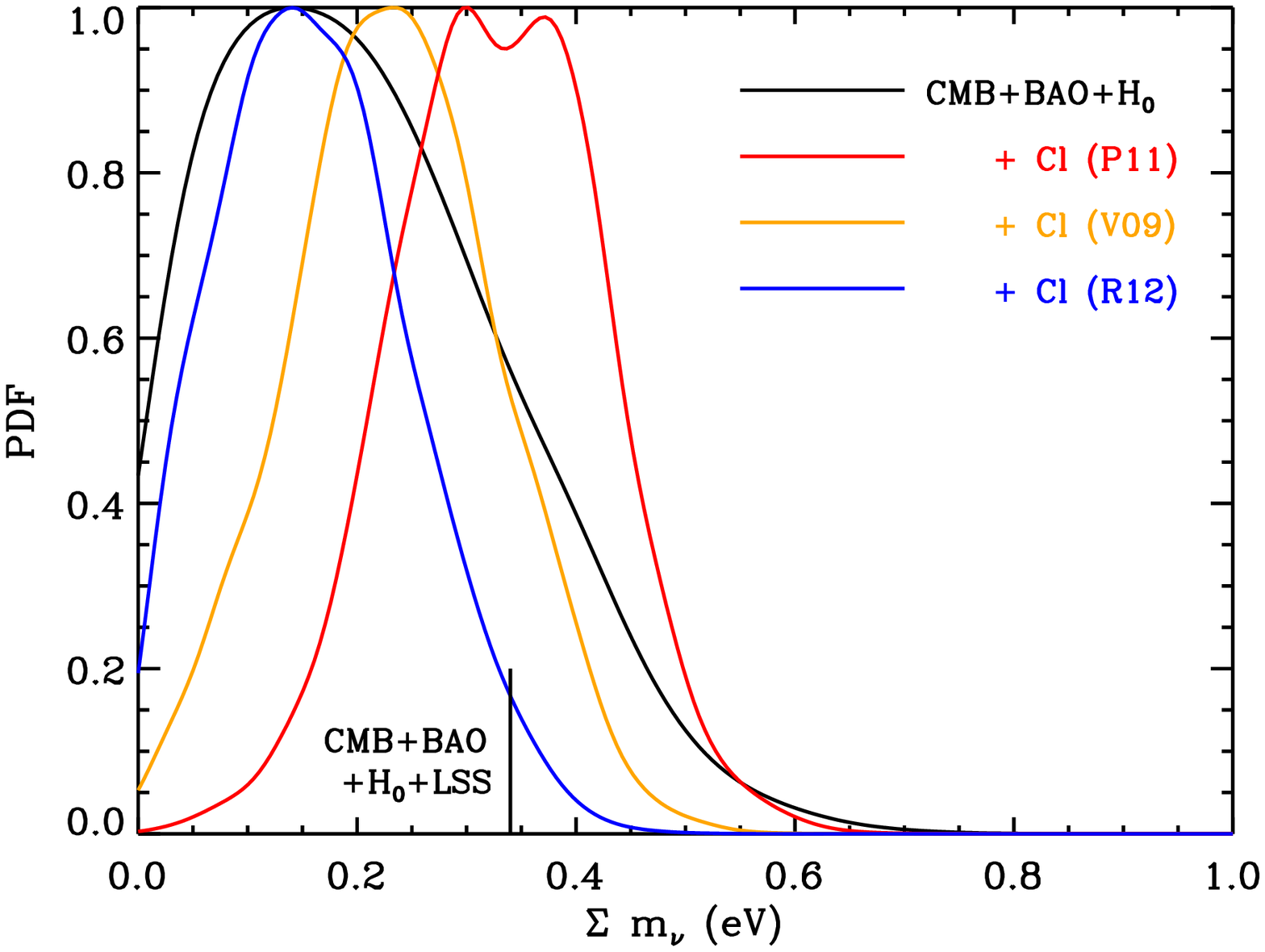}
\caption{{\it Left panel: } Posterior for the sum of neutrino masses $\sum m_\nu$ for a variety
of different analyses. 
The black curve is the WMAP9+BAO result from
\citet{wmap9}, while the red, purple, and blue curves show the posteriors
after inclusion of the cluster normalization condition appropriate for
the \citet{planck11_xray,planck11_local}, \citet{vikhlininetal09}, and
\citet{rozoetal12d} mass calibrations.
{\it Right panel: } As left panel, but starting with CMB+BAO+$H_0$
constraints.  Here, CMB refers to WMAP9+SPT+ACT data.  The small vertical
line at $\sum m_\nu=0.34$ marks the 95\% confidence upper limit derived using
the galaxy correlation function from WiggleZ \citep{parkinsonetal12} and BOSS \citep{zhaoetal12}.
}
\label{fig:lkhd1d}
\vspace{-3mm}
\end{figure*}
\vspace{-2mm}


The left panel in Figure~\ref{fig:lkhd1d} shows the posterior distribution for $\sum m_\nu$ for a combined WMAP9+BAO+Cl
analysis using a variety of different mass calibrations.   In addition to the original V09
mass calibration (orange line), we consider the mass calibration of \citet{arnaudetal10} as employed in
\citet[P11][red curve]{planck11_xray,planck11_local}, and 
that advocated for in \citet[][blue curve, R12]{rozoetal12d}.  
The mass offset $\Delta \ln M$ of the various mass calibrations is defined
as the difference in the log-mass between the various works, averaged over all galaxy
clusters.  So, for instance, the mass shift $\Delta \ln M$ between P11 and V09 was computed by
selecting all galaxy clusters in common to these two cluster samples, computing the difference in the log-mass, and then 
averaging over all such clusters \citep[see][for details]{rozoetal12b}.  A similar analysis for the R12 mass calibration
is done in \citet{rozoetal12d}.  The mass offsets and the corresponding cluster normalization conditions are summarized
in Table \ref{tab:s8}.


\begin{table}
\caption{Mass calibration offset $\Delta \ln M$ relative to \citet[V09][]{vikhlininetal09} for the cluster mass
scale employed in \citet[P11][]{planck11_local} and \citet[R12][]{rozoetal12d}, and the corresponding
cluster mass normalization condition $s_8$.
}
\begin{center}
\begin{tabular}{lcc}
\hline
Reference & $\Delta \ln M$ & $s_8=\sigma_8(\Omega_m/0.25)^\gamma$ \\
\hline
\hline
V09 & --- & 0.813 \\
P11 & $-0.12\pm 0.02$ & 0.781 \\
R12 & $0.11\pm 0.04$ & 0.842
\end{tabular}
\label{tab:s8}
\end{center}
\end{table}


The corresponding constraints on the sum of the neutrino masses for these mass calibrations are
\be
\sum m_\nu = 0.39\pm 0.10\ \eV\ \hspace{0.2 in} \mbox{WMAP9+BAO+Cl (P11)}
\ee
and
\be
\sum m_\nu = 0.21\pm 0.11\ \eV\  \hspace{0.2 in} \mbox{WMAP9+BAO+Cl (R12)}.
\ee

Alternatively, we can also directly determine whether the value of $s_8$ predicted from the combination of WMAP9+BAO
data in a minimal $\Lambda$CDM model ($\sum m_\nu=0$), and compare it to the observational constraints from galaxy
clusters as a test of this minimal cosmological model.   
We find that the \citet[][]{planck11_xray,planck11_local} mass calibration rules out the minimal
flat $\Lambda$CDM model at $3.7\sigma$, whereas the \citet[R12,][]{rozoetal12d} mass calibration
is consistent with a minimal $\Lambda$CDM model at the $1.7\sigma$ level.   Both
results were expected a priori based on the discussion in \citet{rozoetal12d}, where
we showed that the
X-ray luminosity function and the \citet{planck11_xray,planck11_local} mass calibration were inconsistent with WMAP7+BAO
data for a minimal $\Lambda$CDM model. 
Note, however, that the abundance tests in \citet{rozoetal12d} were performed as an {\it a posteriori}
check on the proposed solution to the original Planck--maxBCG discrepancy, i.e. this mass calibration
was not tuned by fitting X-ray and optical abundance data.


\section{The Importance of Cluster Multi-Wavelength Modeling}
\label{sec:power}

While mass calibration suffers from significant uncertainties \citep[see][for a comparison of
cluster masses from the literature]{rozoetal12b,applegateetal12}, we note that multi-wavelength
data can provide compelling evidence for favoring one set of mass calibrations over another.  
In particular, models that are inconsistent with any one aspect of a full
multi-wavelength analysis should be rejected, unless additional presently-unknown systematics 
are identified.

Consider, for instance, 
the Planck--maxBCG discrepancy.  \citet{planck11_optical} noted the observed
SZ signal of maxBCG galaxy clusters was lower than predicted.
Since, \citet{sehgaletal12} found ACT data to be in conflict with both the predicted SZ signal of maxBCG
clusters, as well as the Planck measurements.  
\citet{sehgaletal12} finds that in order for cluster miscentering to account for these offsets one
would require all the maxBCG clusters to be miscentered, with a uniform miscentering kernel extending
out to  $R=1.5\ \Mpc$.
However, this miscentering model is clearly ruled out by X-ray data \citep[see Fig. 2 in][]{sehgaletal12}, which
shows that the central galaxy of maxBCG clusters is most often coincident with the X-ray 
peak \citep[see also ][ and references therein]{menanteauetal13,vonderlindenetal12,mahdavietal12,songetal12,stottetal12}.
Moreover, a miscentering kernel this large would also have a dramatic impact on the weak lensing masses,
further heightening the tension between Planck/ACT and maxBCG.  We concur with \citet{sehgaletal12} that cluster
miscentering cannot be the main explanation for the observed SZ offsets.

As a second example, \citet{planck11_optical} and \citet{anguloetal12} both noted the predicted SZ signal for 
maxBCG galaxy clusters obtained from the path $N_{200} \rightarrow L_X \rightarrow \Ysz$ differed from the prediction 
obtained through the  path $N_{200} \rightarrow M \rightarrow \Ysz$.   That is, the full
set of X-ray, SZ, and optical cluster scaling relations that led to the Planck--maxBCG discrepancy was not internally self-consistent,
a clear signal of systematic errors.  A correct and fully self-consistent set of scaling relation must allow one to go from from any two 
scaling relations to a third and still arrive at the same predicted scaling relation.

As was demonstrated in \citet{rozoetal12d}, lowering the optical mass estimates within its systematic errors while simultaneously increasing the
X-ray mass estimates within its systematic errors results in an overall mass calibration 
that resolves the Planck--maxBCG discrepancy, fits all available X-ray and SZ data, and results
in a self-consistent set of scaling relations.
Moreover, this correctly reproduces optical and X-ray cluster abundances 
in a minimal $\Lambda$CDM model with WMAP7+BAO priors, it correctly predicts the
thermal SZ power spectrum amplitude \citep[e.g.[]{reichardtetal11}, and the resulting masses
are consistent with all
published CLASH data available to date \citep{coeetal12,umetsuetal12}.   
Consequently, we believe there is strong motivation to prefer the mass
calibration advocated in \citet{rozoetal12d}.
We now extend the implications of this mass normalization to a more general cosmological context.


\section{The Role of External Data}

We consider the impact of adding small scale CMB data from  SPT and ACT and external constraints on $H_0$
on our analysis.  We again rely on the
WMAP9 chains and importance sampling, and note that \citet{wmap9} have verified the internal consistency
of these additional external data sets.  The $H_0$ data is that of \citet{riessetal11}, while the SPT and ACT data
are from \citet{keisleretal11} and \citet{dasetal11} respectively.

The right panel in Figure~\ref{fig:lkhd1d} shows the posterior on the sum of neutrino masses derived from the combination 
of CMB+BAO+$H_0$ with galaxy clusters, where CMB includes WMAP9, SPT, and ACT data.
The three colored curves correspond to the three mass normalization conditions we
have discussed in this work: \citet[P11- red curve,][]{planck11_xray,planck11_local}, \citet[V09- orange curve,][]{vikhlininetal09}, 
and \citet[R12- blue curve,][]{rozoetal12d}.
Relative to the left panel, which does not include $H_0$ or small scale CMB  data, 
all posteriors shift to the left, towards lower neutrino masses, and consistent with a higher mass calibration.   
When using this full data set, the 95\% confidence contour obtained with the \citet{rozoetal12d} mass calibration
includes the point $\sum m_\nu=0$, and the corresponding upper limit is
\be
\sum m_\nu \leq 0.32\ \eV\ \hspace{0.2in} \mbox{CMB+BAO+$H_0$+Cl (R12)}.
\ee
Further adding supernovae data as in \citet{wmap9} has a modest impact on our results, 
slightly shifting the neutrino masses further towards $\sum m_\nu=0$.

We can compare our results with those from galaxy correlation function measurements.  
The two most relevant constraints with spectroscopic galaxy samples are those from the WiggleZ \citep{parkinsonetal12} and
BOSS experiments \citep{zhaoetal12}, both of which find $\sum m_\nu \leq 0.34\ \eV$ at the 95\% CL.  Thus, the observed
galaxy correlation function also favors low neutrino masses and a high cluster normalization condition.  We note the main
systematic in these analyses is the impact of non-linearities in the matter power spectrum
and the model for galaxy bias, both of which are expected to be well controlled,
and are completely independent of mass calibration systematics.

We now turn to the recent SPT and ACT results.
Starting with ACT, 
their combined WMAP7+ACT+BAO analysis does not lead
to a detection of neutrino masses \citep{sieversetal13}.  Adding ACT cluster
abundances relying on the dynamical mass calibration of \citet{sifonetal12}
is also consistent with zero neutrino mass \citep{hasselfieldetal13}.
From Table 1 in this last work, we conclude the
\citep{sifonetal12}
mass calibration is $\approx 17\%$ higher than that in \citep{planck11_xray,planck11_local}.
Their posterior from the cosmological analysis is higher still, $\approx 27\%$.  By comparison,
the \citet{rozoetal12d} mass calibration is $\approx 21\%$ higher than that in \citet{planck11_xray,planck11_local}.
Thus, it is not surprising that the ACT cluster analysis does not lead to a detection of neutrino masses.\footnote{It 
is worth nothing that unlike the SPT analysis, ACT does not see any evidence of neutrino
mass from a CMB+BAO only analysis--- compare Fig. 8 in \citet{sieversetal13} with Fig. 9 in \citet{houetal12}. 
Given that the two data sets are consistent \citep{calabreseetal13}, this suggests there may be a 
subtle difference between the two analyses pipelines.}

Turning to SPT, \citet{houetal12} quote a $3\sigma$ detection of neutrino mass, with $\sum m_\nu = 0.32\pm 0.11\ \eV$,
including galaxy clusters.  The cluster constraint from SPT relies on the V09 cluster mass normalization
\citep{bensonetal11}, for which we found $\sum m_\nu = 0.31^{+0.10}_{-0.11}\ \eV$ using WMAP9+BAO+clusters,
in perfect agreement with the SPT result.  In this sense our analysis are consistent, and lowering the mass calibration
of SPT clusters should lead to a reduction of the recovered neutrino masses.

There is, however, one aspect of the \citet{houetal12} results that may appear to contradict the discussion in this
work.  Specifically, 
\citet{houetal12} marginalize over the systematic uncertainty in mass calibration quoted in V09, finding that the posterior
in the neutrino masses is essentially independent of the width of this prior, and contrary to what one might expect
given our discussion.
There is, however, no inconsistency.  This is best understood using Figure \ref{fig:lkhd2d}.  Clusters provide a tight
constraint on the cluster normalization condition $s_8$, which is degenerate with $\sum m_\nu$ in the CMB+BAO+$H_0$
data set.   As long as the constraint on $s_8$ is modest, clusters simply ``pick out'' an $s_8$ value, and the posterior
on the neutrino mass only reflects the CMB+BAO+$H_0$ data with an effective infinitely sharp prior on $s_8$, so
the posterior on $\sum m_\nu$ is insensitive to the precise width of the $s_8$ prior (i.e. the mass calibration uncertainty).
Nevertheless, shifting the central value $s_8$
still slides the cosmological constraints along the $\sum m_\nu$--$s_8$ degeneracy.
In short, the recovered neutrino masses are robust to changes in the width of the $s_8$ (or mass calibration) prior,
but not to shifts in the corresponding central value.


\section{On the CMB and BAO+$H_0$ Tension}


\begin{figure}
\hspace{-5mm} 
\includegraphics[width=90mm]{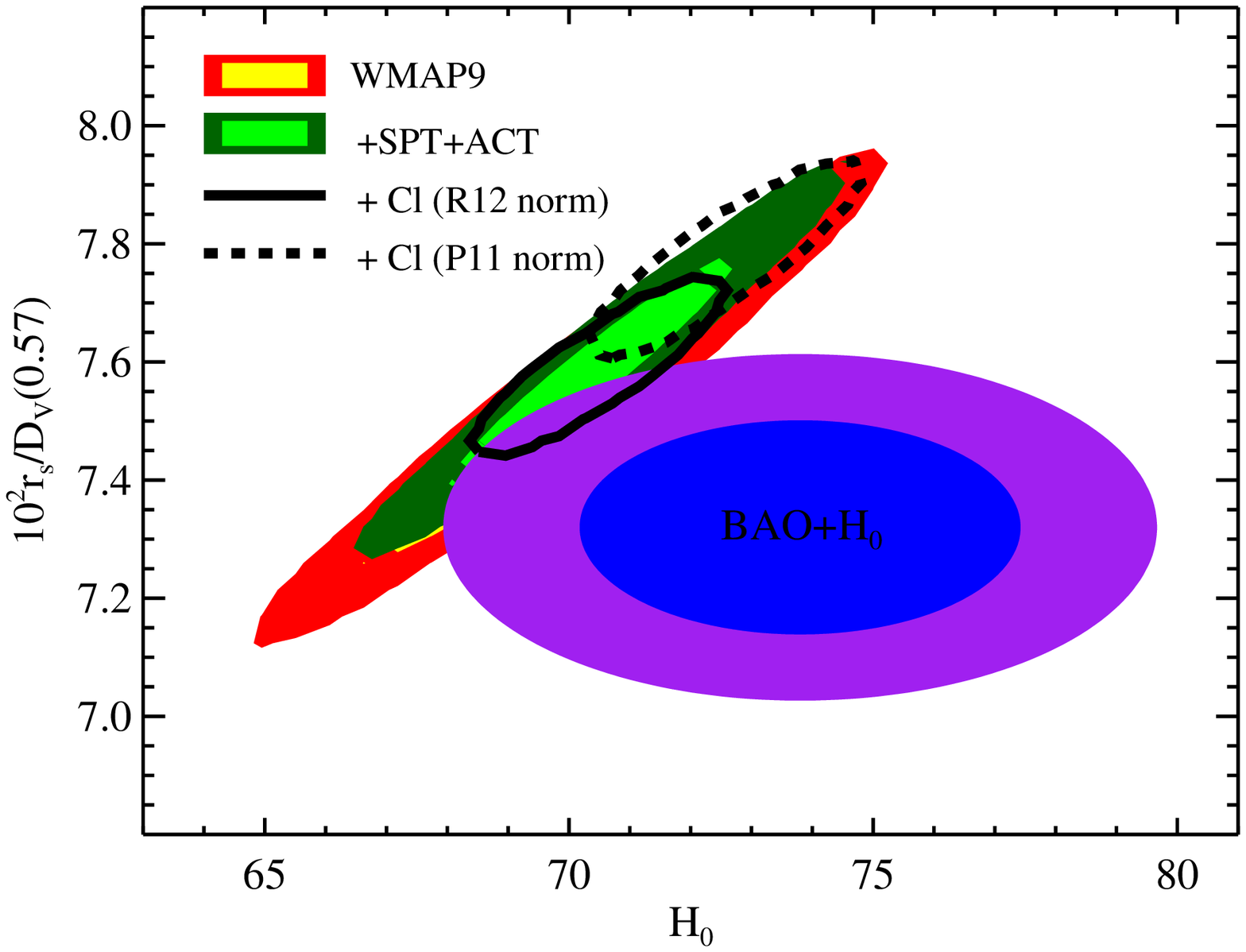}
\caption{Comparison of the 68\% and 95\% likelihood contours 
in the  $r_s/D_V(0.57)$--$H_0$ plane from
various data sets assuming a minimal $\Lambda$CDM model, as labelled.  
The modest tension between CMB and BAO+$H_0$ data 
increases when adding galaxy clusters with a low cluster mass
normalization \citep[e.g.][]{planck11_xray,planck11_local}.
The \citet{rozoetal12d} mass normalization shrinks the CMB
contours around the CMB and BAO+$H_0$ overlap region.
}
\vspace{-3mm}
\label{fig:tension}
\end{figure}


As emphasized by \citet{houetal12}, one critical driving force behind the SPT detection of neutrino
masses is the mild tension ($\leq 2\sigma$) between CMB and BAO+$H_0$ data in a minimal $\Lambda$CDM cosmology.  
We follow \citet{houetal12} and illustrate
this tension in Figure~\ref{fig:tension} by comparing current CMB constraints on 
$r_s/D_V(0.57)$ and $H_0$
for a minimal $\Lambda$CDM model  to the BAO measurements of \citet{andersonetal12} and $H_0$ measurements
from \citet{riessetal11}.   Here, $r_s$ is the comoving sound horizon at decoupling, and
\be
D_V(z) = \left [z(1+z)^2D_A^2(z)cH^{-1}(z) \right]^{1/3}
\ee
where $D_A$ is the angular diameter distance.

Consider now how galaxy clusters affect this discussion.    The solid and dashed ellipses in Figure~\ref{fig:tension} show the $95\%$ confidence
contours obtained when adding a cluster normalization condition prior to the CMB data, using the \citet{rozoetal12d} and 
\citet{planck11_xray,planck11_local} mass calibrations respectively.  Galaxy clusters tighten the confidence regions in the
$r_s/D_V(0.57)$--$H_0$ plane, but that the choice of mass calibration shifts the posterior for the combined data sets 
along the CMB degeneracy curve.  Of the 3 mass calibration we consider here,  
the \citet{rozoetal12d} mass calibration falls closest to the intersection of the CMB and BAO+$H_0$ data;
In such a scenario, the current ``tension'' could easily reflect a statistical fluctuation, 
in which case the various contours
will likely shift towards each other as the uncertainties in the measurements continue to decrease.

By contrast, a low mass calibration such as that 
of \citet{planck11_xray,planck11_local} heightens the existing 
tension between CMB and BAO+$H_0$ data. Moreover, while 
non-zero neutrino masses can reconcile this data with galaxy clusters, this new degree of freedom does not help
alleviate the tension between the CMB and BAO+$H_0$ data sets.
Indeed, this tension is most effectively alleviated 
by increasing the effective number of neutrino species
\citep[not shown, but see Figure 3 in][]{houetal12}.
Since in this scenario our minimal $\Lambda$CDM model is already ruled out, 
there is really nothing ``pulling'' the BAO+$H_0$ and CMB contours towards better agreement, in which
case the tension between these data sets is likely to increase as the error bars decrease.
If so, we should expect the evidence for massive neutrinos and a non-standard number of neutrino
species to become compelling in the near future.  Indeed, \citet{burenin13} --- a paper that appeared
on the archive as we were finishing this work --- argues
for such a detection on the basis of current CMB+BAO+$H_0$ and cluster data from V09.


\section{Summary and Discussion}
\label{sec:summary}

Galaxy clusters are the key piece of data required
to place tight cosmological constraints on the sum of the neutrino masses, with the resulting constraints being critically
sensitive to the adopted mass calibration.   We have noted that the self-consistency
of optical, X-ray, and SZ scaling relations favor a high mass calibration, which 
helps reconcile CMB and BAO+$H_0$ data in a minimal $\Lambda$CDM cosmological model,
and allows us to place an upper limit on the sum of the neutrino masses, $\sum m_\nu \leq 0.32$ (95\% CL).   
This constraint is nearly identical to that derived from CMB+BAO+$H_0$ data combined with measurements of
the galaxy power spectrum.
By contrast, a lower mass calibration rules out the minimal $\Lambda$CDM model, and necessitates a non-zero neutrino mass.

This picture leaves us with an interesting dichotomy: the \citet{rozoetal12d} mass calibration 
suggests that the slight tension in the CMB and BAO+$H_0$ data sets will 
likely decrease as error bars decrease.
Conversely, should the low mass calibration of \citet{planck11_xray,planck11_local} be correct, the tension in the CMB and BAO+$H_0$
data is at least in part due to a breakdown of the minimal $\Lambda$CDM model.  
In that case, this tension
is likely to increase with decreasing errors, requiring both a non-zero neutrino mass and a non-minimal effective number of neutrino species.
The \citet{vikhlininetal09} mass calibration falls somewhere in between, showing evidence for massive neutrinos at $2.7\sigma$, and
also favoring a high $N_{eff}$ \citep{burenin13}.

In short, it is clear that the combination of improved BOSS constraints and Planck data will provide an important test 
of the various cluster mass normalizations advocated for in the literature.  From the point of view of internal self-consistency
of X-ray, SZ, and optical data, we note that should the mass calibration advocated in \citet{rozoetal12d} be ruled out, 
then the existing X-ray, SZ, and optical data will be in tension once again.  In this case, the resolution of the tension will likely require
the identification of an additional, currently-unknown systematic in at least one of these data sets. Note too that because
current X-ray mass calibrations assume no hydrostatic bias, it will also become important to understand why hydrodynamical
simulations generically predict $\approx 10\%-30\%$
hydrostatic biases \citep[e.g.][]{nagaietal07a,lauetal09,battagliaetal11,nelsonetal11,rasiaetal12,sutoetal13}. 
Whatever the case may be, we find it beautifully ironic that 
the key systematic in our ability to weigh $\approx 0.1\ \eV$ neutrinos is our ability to weigh $\approx 10^{14}-10^{15}\ \msun$ objects.


\section*{Acknowledgments} 

E.R. would like to thank Ryan Keisler, Bradford Benson, Christian Reichardt, and Tijmen de Haan for help
in understanding the details of the recent analysis of \citet{houetal12}, and Raul Angulo and Joanne Cohn
for comments on an earlier version of this manuscript.  The authors thank David Weinberg and Matthew
Becker for a careful reading of this work, and many insightful comments.
This work was supported in part by  the U.S. Department of Energy contract to SLAC no. DE-AC02-76SF00515.
AEE acknowledges support from NASA NNX07AN58G.  
JGB gratefully acknowledges support from the Institut Universitaire
de France. A portion of the research described
in this paper was carried out at the Jet Propulsion Laboratory,
California Institute of Technology, under a contract
with the National Aeronautics and Space Administration.

\newcommand\AAA{{A\& A}}
\newcommand\PhysRep{{Physics Reports}}
\newcommand\apj{{ApJ}}
\newcommand\PhysRevD[3]{ {Phys. Rev. D}} 
\newcommand\prd[3]{ {Phys. Rev. D}} 
\newcommand\PhysRevLet[3]{ {Phys. Rev. Letters} }
\newcommand\mnras{{MNRAS}}
\newcommand\PhysLet{{Physics Letters}}
\newcommand\AJ{{AJ}}
\newcommand\aap{ {A \& A}}
\newcommand\apjl{{ApJ Letters}}
\newcommand\aph{astro-ph/}
\newcommand\AREVAA{{Ann. Rev. A.\& A.}}
\newcommand\pasj{PASJ}
\newcommand\apjs{{ApJ Supplement}}
\newcommand\jcap{JCAP}

\bibliographystyle{mn2e}
\bibliography{mybib}

\begin{thebibliography}{}

\bibitem[\protect\citeauthoryear{{Allen} et~al.,}{{Allen}
  et~al.}{2011}]{allenetal11}
{Allen} S.~W.,  et~al., 2011, ArXiv:1103.4829

\bibitem[\protect\citeauthoryear{{Anderson} et~al.,}{{Anderson}
  et~al.}{2012}]{andersonetal12}
{Anderson} L.,  et~al., 2012, \mnras, 427, 3435

\bibitem[\protect\citeauthoryear{{Angulo} et~al.,}{{Angulo}
  et~al.}{2012}]{anguloetal12}
{Angulo} R.~E.,  et~al., 2012, ArXiv:1203.3216

\bibitem[\protect\citeauthoryear{{Applegate} et~al.,}{{Applegate}
  et~al.}{2012}]{applegateetal12}
{Applegate} D.~E.,  et~al., 2012, ArXiv: 1208.0605

\bibitem[\protect\citeauthoryear{{Arnaud} et~al.,}{{Arnaud}
  et~al.}{2010}]{arnaudetal10}
{Arnaud} M.,  et~al., 2010, \aap, 517, A92+

\bibitem[\protect\citeauthoryear{{Battaglia} et~al.,}{{Battaglia}
  et~al.}{2011}]{battagliaetal11}
{Battaglia} N.,  et~al., 2011, ArXiv: 1109.3709

\bibitem[\protect\citeauthoryear{{Benson} et~al.,}{{Benson}
  et~al.}{2011}]{bensonetal11}
{Benson} B.~A.,  et~al., 2011, ArXiv:1112.5435

\bibitem[\protect\citeauthoryear{{Beutler} et~al.,}{{Beutler}
  et~al.}{2011}]{beutleretal11}
{Beutler} F.,  et~al., 2011, \mnras, 416, 3017

\bibitem[\protect\citeauthoryear{{Blake} et~al.,}{{Blake}
  et~al.}{2012}]{blakeetal12}
{Blake} C.,  et~al., 2012, \mnras, 425, 405

\bibitem[\protect\citeauthoryear{{Burenin}}{{Burenin}}{2013}]{burenin13}
{Burenin} R.~A.,  2013, ArXiv: 1301.4791

\bibitem[\protect\citeauthoryear{{Burenin} \& {Vikhlinin}}{{Burenin} \&
  {Vikhlinin}}{2012}]{bureninvikhlinin12}
{Burenin} R.~A.,  {Vikhlinin} A.~A.,  2012, Astronomy Letters, 38, 347

\bibitem[\protect\citeauthoryear{{Calabrese} et~al.,}{{Calabrese}
  et~al.}{2013}]{calabreseetal13}
{Calabrese} E.,  et~al., 2013, ArXiv e-prints

\bibitem[\protect\citeauthoryear{{Coe} et~al.,}{{Coe}
  et~al.}{2012}]{coeetal12}
{Coe} D.,  et~al., 2012, ArXiv:1201.1616

\bibitem[\protect\citeauthoryear{{Das} et~al.,}{{Das}
  et~al.}{2011}]{dasetal11}
{Das} S.,  et~al., 2011, \apj, 729, 62

\bibitem[\protect\citeauthoryear{{Hasselfield} et~al.,}{{Hasselfield}
  et~al.}{2013}]{hasselfieldetal13}
{Hasselfield} M.,  et~al., 2013, ArXiv: 1301.0816

\bibitem[\protect\citeauthoryear{{Hinshaw} et~al.,}{{Hinshaw}
  et~al.}{2012}]{wmap9}
{Hinshaw} G.,  et~al., 2012, ArXiv: 1212.5226

\bibitem[\protect\citeauthoryear{{Hou} et~al.,}{{Hou}
  et~al.}{2012}]{houetal12}
{Hou} Z.,  et~al., 2012, ArXiv: 1212.6267

\bibitem[\protect\citeauthoryear{{Keisler} et~al.,}{{Keisler}
  et~al.}{2011}]{keisleretal11}
{Keisler} R.,  et~al., 2011, \apj, 743, 28

\bibitem[\protect\citeauthoryear{{Lau}, {Kravtsov} \& {Nagai}}{{Lau}
  et~al.}{2009}]{lauetal09}
{Lau} E.~T.,  {Kravtsov} A.~V.,    {Nagai} D.,  2009, \apj, 705, 1129

\bibitem[\protect\citeauthoryear{{Lesgourgues} \& {Pastor}}{{Lesgourgues} \&
  {Pastor}}{2012}]{lesgourguespastor12}
{Lesgourgues} J.,  {Pastor} S.,  2012, ArXiv: 1212.6154

\bibitem[\protect\citeauthoryear{{Mahdavi} et~al.,}{{Mahdavi}
  et~al.}{2012}]{mahdavietal12}
{Mahdavi} A.,  et~al., 2012, ArXiv: 1210.3689

\bibitem[\protect\citeauthoryear{{Mantz} et~al.,}{{Mantz}
  et~al.}{2010}]{mantzetal10c}
{Mantz} A.,  et~al., 2010, \mnras, 406, 1805

\bibitem[\protect\citeauthoryear{{Menanteau} et~al.,}{{Menanteau}
  et~al.}{2013}]{menanteauetal13}
{Menanteau} F.,  et~al., 2013, \apj, 765, 67

\bibitem[\protect\citeauthoryear{{Nagai}, {Vikhlinin} \& {Kravtsov}}{{Nagai}
  et~al.}{2007}]{nagaietal07a}
{Nagai} D.,  {Vikhlinin} A.,    {Kravtsov} A.~V.,  2007, \apj, 655, 98

\bibitem[\protect\citeauthoryear{{Nelson} et~al.,}{{Nelson}
  et~al.}{2011}]{nelsonetal11}
{Nelson} K.,  et~al., 2011, ArXiv: 1112.3659

\bibitem[\protect\citeauthoryear{{Padmanabhan} et~al.,}{{Padmanabhan}
  et~al.}{2012}]{padmanabhanetal12}
{Padmanabhan} N.,  et~al., 2012, \mnras, 427, 2132

\bibitem[\protect\citeauthoryear{{Parkinson} et~al.,}{{Parkinson}
  et~al.}{2012}]{parkinsonetal12}
{Parkinson} D.,  et~al., 2012, \prd, 86, 103518

\bibitem[\protect\citeauthoryear{{Planck Collaboration}}{{Planck
  Collaboration}}{2011a}]{planck11_xray}
{Planck Collaboration} 2011a, \aap, 536, A10

\bibitem[\protect\citeauthoryear{{Planck Collaboration}}{{Planck
  Collaboration}}{2011b}]{planck11_local}
{Planck Collaboration} 2011b, \aap, 536, A11

\bibitem[\protect\citeauthoryear{{Planck Collaboration}}{{Planck
  Collaboration}}{2011c}]{planck11_optical}
{Planck Collaboration} 2011c, \aap, 536, A12

\bibitem[\protect\citeauthoryear{{Rasia} et~al.,}{{Rasia}
  et~al.}{2012}]{rasiaetal12}
{Rasia} E.,  et~al., 2012, ArXiv:1201.1569

\bibitem[\protect\citeauthoryear{{Reichardt} et~al.,}{{Reichardt}
  et~al.}{2011}]{reichardtetal11}
{Reichardt} C.~L.,  et~al., 2011, ArXiv:1111.0932

\bibitem[\protect\citeauthoryear{{Reid} et~al.,}{{Reid}
  et~al.}{2010}]{reidetal10}
{Reid} B.~A.,  et~al., 2010, \jcap, 1, 3

\bibitem[\protect\citeauthoryear{{Riess} et~al.,}{{Riess}
  et~al.}{2011}]{riessetal11}
{Riess} A.~G.,  et~al., 2011, \apj, 730, 119

\bibitem[\protect\citeauthoryear{{Rozo} et~al.,}{{Rozo}
  et~al.}{2012a}]{rozoetal12b}
{Rozo} E.,  et~al., 2012a, ArXiv: 1204.6301

\bibitem[\protect\citeauthoryear{{Rozo} et~al.,}{{Rozo}
  et~al.}{2012b}]{rozoetal12d}
{Rozo} E.,  et~al., 2012b, ArXiv: 1204.6305

\bibitem[\protect\citeauthoryear{{Sehgal} et~al.,}{{Sehgal}
  et~al.}{2012}]{sehgaletal12}
{Sehgal} N.,  et~al., 2012, ArXiv:1205.2369

\bibitem[\protect\citeauthoryear{{Sievers} et~al.,}{{Sievers}
  et~al.}{2013}]{sieversetal13}
{Sievers} J.~L.,  et~al., 2013, ArXiv: 1301.0824

\bibitem[\protect\citeauthoryear{{Sifon} et~al.,}{{Sifon}
  et~al.}{2012}]{sifonetal12}
{Sifon} C.,  et~al., 2012, ArXiv: 1201.0991

\bibitem[\protect\citeauthoryear{{Song} et~al.,}{{Song}
  et~al.}{2012}]{songetal12}
{Song} J.,  et~al., 2012, \apj, 761, 22

\bibitem[\protect\citeauthoryear{{Stott} et~al.,}{{Stott}
  et~al.}{2012}]{stottetal12}
{Stott} J.~P.,  et~al., 2012, \mnras, 422, 2213

\bibitem[\protect\citeauthoryear{{Suto} et~al.,}{{Suto}
  et~al.}{2013}]{sutoetal13}
{Suto} D.,  et~al., 2013, ArXiv: 1302.5172

\bibitem[\protect\citeauthoryear{{Umetsu} et~al.,}{{Umetsu}
  et~al.}{2012}]{umetsuetal12}
{Umetsu} K.,  et~al., 2012, ArXiv:1204.3630

\bibitem[\protect\citeauthoryear{{Vikhlinin} et~al.,}{{Vikhlinin}
  et~al.}{2009a}]{vikhlininetal09}
{Vikhlinin} A.,  et~al., 2009a, \apj, 692, 1033

\bibitem[\protect\citeauthoryear{{Vikhlinin} et~al.,}{{Vikhlinin}
  et~al.}{2009b}]{vikhlininetal09b}
{Vikhlinin} A.,  et~al., 2009b, \apj, 692, 1060

\bibitem[\protect\citeauthoryear{{von der Linden}, {Allen}, {Applegate},
  {Kelly}, {Allen}, {Ebeling}, {Burchat}, {Burke}, {Donovan}, {Morris},
  {Blandford}, {Erben} \& {Mantz}}{{von der Linden}
  et~al.}{2012}]{vonderlindenetal12}
{von der Linden} A.,  {Allen} M.~T.,  {Applegate} D.~E.,  {Kelly} P.~L.,
  {Allen} S.~W.,  {Ebeling} H.,  {Burchat} P.~R.,  {Burke} D.~L.,  {Donovan}
  D.,  {Morris} R.~G.,  {Blandford} R.,  {Erben} T.,    {Mantz} A.,  2012,
  ArXiv: 1208.0597

\bibitem[\protect\citeauthoryear{{Weinberg} et~al.,}{{Weinberg}
  et~al.}{2012}]{weinbergetal12}
{Weinberg} D.~H.,  et~al., 2012, ArXiv:1201.2434

\bibitem[\protect\citeauthoryear{{Zhao} et~al.,}{{Zhao}
  et~al.}{2012}]{zhaoetal12}
{Zhao} G.-B.,  et~al., 2012, ArXiv: 1211.3741

\end{thebibliography}

\appendix



\label{lastpage}

\end{document}